\documentclass[conference]{IEEEtran}
\IEEEoverridecommandlockouts
\usepackage{subfigure}
\usepackage{multirow}
\usepackage{cite}
\usepackage{amsmath,amssymb,amsfonts}
\usepackage{algorithmic}
\usepackage{graphicx}
\usepackage{textcomp}
\usepackage{xcolor}
\def\BibTeX{{\rm B\kern-.05em{\sc i\kern-.025em b}\kern-.08em
    T\kern-.1667em\lower.7ex\hbox{E}\kern-.125emX}}
\begin{document}

\title{\textcolor{red}{AMG:} Automated Efficient \underline{A}pproximate \underline{M}ultiplier \underline{G}enerator for FPGAs via \textcolor{red}{Bayesian Optimization}\\
}

\author{\IEEEauthorblockN{Zhen Li, Hao Zhou, and Lingli Wang\textsuperscript{$\dagger$}}
\IEEEauthorblockA{\textit{State Key Laboratory of ASIC and System}\\
\textit{School of Microelectronics, Fudan University}\\
Shanghai, China\\
\textsuperscript{$\dagger$}llwang@fudan.edu.cn}
}
\maketitle

\begin{abstract}
Approximate computing is a promising approach to reduce the power, delay, and area in hardware design for many error-resilient applications such as machine learning (ML) and digital signal processing (DSP) systems, in which multipliers usually are key arithmetic units. Due to the underlying architectural differences between ASICs and FPGAs, existing ASIC-based approximate multipliers do not offer symmetrical gains when they are implemented by FPGA resources. In this paper, we propose \textcolor{red}{AMG}, an open-source automated approximate multiplier generator for FPGAs driven by \textcolor{red}{Bayesian optimization (BO)} with parallel evaluation. The proposed method simplifies the exact half adders (HAs) for the initial partial product (PP) compression in a multiplier while preserving coarse-grained additions for the following accumulation. The generated multipliers can be effectively mapped to lookup tables (LUTs) and carry chains provided by modern FPGAs, reducing hardware costs with acceptable errors. Compared with 1167 multipliers from previous works, our generated multipliers can form a Pareto front with 28.70\%-38.47\% improvements in terms of the product of hardware cost and error on average. All source codes, reproduced multipliers, and our generated multipliers are available at https://github.com/phyzhenli/AMG.
\end{abstract}


\section{Introduction}

Moore's law, which is named after Gordon Moore, was observed in 1965 that states the number of transistors in an integrated circuit (IC) doubles approximately every two years \cite{Approximate:SurveyMoore}. For decades, semiconductor manufacturers have been following Moore’s prediction and continuously shrinking the size of transistors. Meanwhile, IC designers have developed many optimization methods to achieve high performance by utilizing the continuous supply of surplus transistors.
In 1974, Robert Dennard proposed Dennard Scaling which states that as transistors \textcolor{black}{become} smaller, their power density stays constant. Unfortunately, since about 2007, as the transistors continued to shrink, the leakage current increased, leading to the breakdown of Dennard Scaling.
Approximate computing is a promising approach to solve the rising power efficiency challenges. It is a novel computation paradigm based on the fact that many applications can tolerate errors, which can be used to trade off for power, performance, and area improvements.
Both hardware-level and software-level approximations can be applied for performing approximate computing. In hardware design, many approximate arithmetic circuits such as adders, multipliers, and dividers \textcolor{black}{have been} proposed to reduce the hardware overhead. This paper concentrates on approximate multipliers.

A classic design was proposed in \cite{Approximate:KMap}, in which a 2$\times$2 exact multiplier was manually simplified according to its Karnaugh map to construct large bit-width approximate multipliers.
In \cite{Approximate:Compressor}, two approximate 4-2 compressors and four different schemes were proposed and analyzed for a Dadda multiplier with an extensive simulation. Reference \cite{Approximate:AC} \textcolor{black}{utilized} approximate half adder (HA), full adder (FA), and 4-2 compressor to reduce altered partial products (PPs) produced by OR gates. In \cite{Approximate:CR}, an approximate adder with limited carry propagation was utilized for a fast \textcolor{black}{PP} accumulation, achieving a shorter critical path than traditional multipliers.
In addition to manual methods, mathematical approximations have been studied to design approximate multipliers. The essence of mathematical methods is to replace the exact multiplication with operations that are more efficient for circuit implementation. For example, reference \cite{Approximate:Optimal} uses inversion, shift, and addition to approximate floating-point multiplication. It is mathematically proved that the proposed multiplier is optimal in terms of the squared error based on the given bases of the space $\{1, x, y, x^2, y^2\}$.
Automated design space exploration methods have also been utilized to generate approximate multipliers. One of the most famous methods is based on Cartesian Genetic Programming (CGP). In \cite{Approximate:CGP2016NN}, CGP was first introduced to generate inexact multipliers with given constraints for Artificial Neural Networks. Later, a series of works based on CGP has brought the research on approximate computing to a climax \cite{Approximate:CGPDCT,Approximate:EvoApprox8b,Approximate:EvoLite,Approximate:ALWANN}.

The designs mentioned above are all for ASICs, which means that their hardware implementations are composed of logic gates.
\textcolor{black}{Modern FPGAs provide DSP blocks to achieve fast multiplications. However, DSP blocks consume only 5\% of the FPGA core area in DSP-rich devices with fixed locations \cite{FPGA:DSP}, which means that their usage may not be efficient for some complex applications such as deep neural networks (DNNs) \cite{FPGA:Math}. In contrast, the most abundant resource in an FPGA is the lookup table (LUT), which is capable of implementing different functions with rich routing resources. It is possible to achieve high-performance multiplier designs with the efficient utilization of LUTs.
Unfortunately,} due to the intrinsic architectural differences between ASICs and FPGAs, existing ASIC-oriented approximate multipliers usually provide limited performance gains when they are directly synthesized for FPGAs. There are also some works customized toward FPGA-based fabrics. In \cite{Approximate:SMApproxLib}, the first open-source FPGA-based approximate multipliers library was proposed \textcolor{black}{by} modifying INIT parameter values in LUT primitives with associated carry chains. There are also some \textcolor{black}{papers based on} the similar LUT encoding approach to generate softcore multipliers with different hardware costs and errors \cite{Approximate:CaCc,Approximate:FPT22,Approximate:TCAD22}. However, the manual-design process is time-consuming and the generated approximate multipliers usually have significant performance differences.
Thus there is a demand for a fully automated design of approximate multipliers specifically for FPGAs.

The main contributions of this paper are as follows:
\begin{itemize}
    \item An automated approach for generating the HA array which is used for the initial \textcolor{black}{PP} compression in a multiplier is proposed to support various bit widths. Meanwhile, we set a HA's weight \textcolor{black}{based on the binary weight of its two inputs, which reflects the HA's significance to the multiplier output.}
    \item We design an efficient simplification method for an exact HA to construct a large search space. Based on weights, the number of HAs to be optimized can be adjusted according to the desired area (i.e. LUT count) percentage reduction.
    \item We propose a platform to automatically search high-performance approximate multipliers via \textcolor{red}{Bayesian optimization (BO)} with parallel evaluation. The PPs compressed by searched HAs are then summed up by coarse-grained additions directly. The straightforward additions can be efficiently recognized by electronic design automation (EDA) tools (e.g. Vivado) and mapped to fast carry chains.
\end{itemize}

The rest of the paper is organized as follows. Section \ref{Sec:backgroud} introduces the background of our work. Section \ref{Sec:design} describes the proposed design for automated approximate multiplier generation. Section \ref{Sec:results} presents the experimental results of our generated approximate multipliers compared with \textcolor{black}{many} multipliers from previous works. Finally, Section \ref{Sec:conclusion}  summarizes and concludes this paper.

\section{Background}\label{Sec:backgroud}

\subsection{\textcolor{black}{Comparison Between ASICs \& FPGAs for Multipliers}}\label{SubSec:PDA}

Due to the inherent architectural differences between ASICs and FPGAs, approximate multipliers designed for ASICs usually cannot achieve comparable performance gains for FPGAs. To illustrate the effect, we compare the product of power, delay, and area (PDA) percentage improvements of EvoApprox8b \cite{Approximate:EvoApprox8b} multipliers based on FPGAs and ASICs. In ASIC design, we can easily obtain the silicon area of a chip. However, in FPGA design, we usually measure the size of a circuit by its occupied resources. Thus we apply the number of used LUTs \textcolor{black}{for} the multiplier \textcolor{black}{as} its FPGA-based implementation.

\begin{figure}[!ht]
    \centering
    \includegraphics[width=0.9\linewidth]{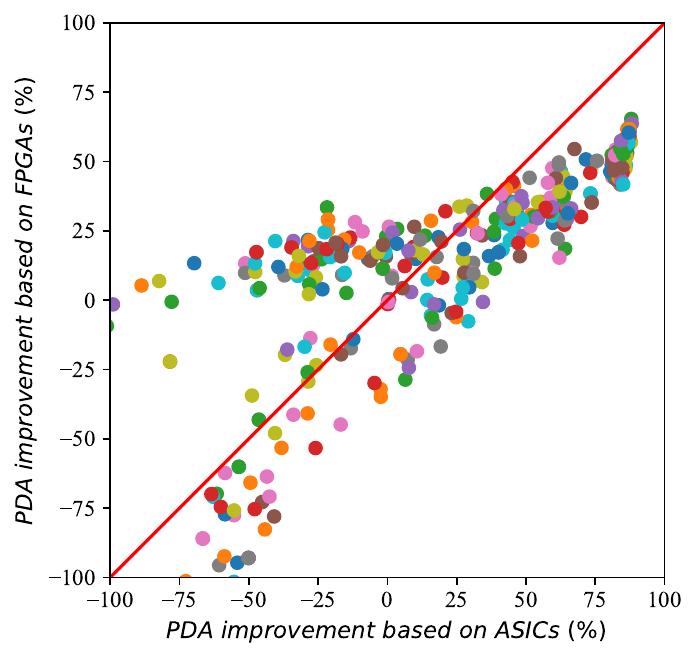}
    \caption{PDA percentage improvements comparison of EvoApprox8b \cite{Approximate:EvoApprox8b} multipliers based on ASICs and FPGAs.}
    \label{Fig:EvoApprox8b_ASIC_FPGA}
\end{figure}
Fig.~\ref{Fig:EvoApprox8b_ASIC_FPGA} shows the comparison of the PDA percentage improvements of EvoApprox8b \cite{Approximate:EvoApprox8b} multipliers based on ASICs and FPGAs. 
The ASIC-based PDA results are reported by Design Compiler (DC) vL-2016.03-SP1 after synthesis under 2GHz clock frequency constraint with Arizona State Predictive PDK (ASAP) 7nm process library \cite{Approximate:ASAP7}. The FPGA-based PDA results are obtained from Vivado 2023.1 after place and routing under 100MHz clock frequency constraint with a Virtex UltraScale+ device xcvu3p-ffvc1517-3-e.
The PDA percentage improvement of an ASIC- or FPGA-based multiplier is defined as:
\begin{equation}
    \label{Eq:PDA_Imp}
    \frac{\text{PDA}_{ext} - \text{PDA}_{app}}{\text{PDA}_{ext}} *100
\end{equation}
where $\text{PDA}_{app}$ and $\text{PDA}_{ext}$ are two PDA values of the approximate multiplier and the exact multiplier respectively. The latter is implemented using a Verilog ``$*$'' operator, which is usually built from Synopsys DesignWare Library \cite{IP:DesignWare} in DC or Xilinx Multiplier LogiCORE IP \cite{IP:LOGICORE} in Vivado.
In Fig.~\ref{Fig:EvoApprox8b_ASIC_FPGA}, each dot represents a multiplier with two PDA percentage improvements by \eqref{Eq:PDA_Imp}. Points on the red line indicate that the corresponding multipliers offer \textcolor{black}{equivalent} hardware cost gains for ASICs and FPGAs. We can see that most of these ASIC-oriented approximate multipliers cannot provide similar performance gains when they are implemented by FPGAs. There are even some approximate multipliers that have 25\% \textcolor{black}{worse} PDA than the exact multiplier based on ASICs but achieve 25\% improvement based on FPGAs. This is because EvoApprox8b \cite{Approximate:EvoApprox8b} multipliers are constructed by 2- or 3-input logic gates, which can be fully controlled over resource utilization at a fine granularity in ASIC design. However, FPGAs use configurable LUTs as logic blocks, which have a completely different property compared to ASIC gates. Besides, in FPGAs, routing accounts for about 50\% of the critical path delay \cite{FPGA:Vaughn}. 
As a result, we need to design LUT-based approximate multipliers for FPGA-based systems to achieve significant performance improvements.

\subsection{Error Metrics}

There are many error metrics developed to evaluate the error of approximate multipliers \cite{Approximate:EvoApprox8b}. In this paper, we mainly consider the mean absolute error (MAE) and the mean squared error (MSE) with the unsigned integer multiplier, which can be defined as:
\begin{equation}
    \label{Eq:MAE}
    \begin{aligned}
        \text{MAE} = \sum_{x=0}^{2^N-1} \sum_{y=0}^{2^M-1} |D(x,y)| \times p(x^{\prime}, y^{\prime})
    \end{aligned}
\end{equation}
\begin{equation}
    \label{Eq:MSE}
    \begin{aligned}
        \text{MSE} = \sum_{x=0}^{2^N-1} \sum_{y=0}^{2^M-1} |D(x,y)|^2 \times p(x^{\prime}, y^{\prime})
    \end{aligned}
\end{equation}
where 
\begin{equation}
    \label{Eq:D}
    D(x,y) = O_{app}(x,y) - O_{ext}(x,y)
\end{equation}
\begin{equation}
    \label{Eq:Pxy}
    p(x^{\prime}, y^{\prime}) = \left\{
        \begin{aligned}
        p_1(x) \times p_2(y),\ \ x^{\prime}=x\ and\ y^{\prime}=y. \\
        p_1(y) \times p_2(x),\ \ x^{\prime}=y\ and\ y^{\prime}=x.
        \end{aligned}
        \right.
\end{equation}
In \eqref{Eq:MAE}-\eqref{Eq:Pxy}, $x$ and $y$ are two inputs of a multiplier with bit widths of $N$ and $M$ respectively. $O_{app}(x,y)$ and $O_{ext}(x,y)$ are approximate and exact \textcolor{black}{outputs} respectively. $p_1$ and $p_2$ are two probability distributions that can be assumed as uniform or extracted from the application multiplications. Under uniform distributions, $p_1$ and $p_2$ are constants and \textcolor{black}{$p_1 \times p_2 = \cfrac{1}{2^{N+M}}$}, which is consistent with equations presented in \cite{Approximate:EvoApprox8b}. However, under non-uniform distributions, an approximate multiplier has different values of MAE and MSE with different input polarities \cite{Approximate:TVLSI}, which are indicated by \eqref{Eq:Pxy}.
This paper only considers the uniform distribution.

\subsection{\textcolor{red}{Bayesian Optimization (BO)}}

BO \cite{BBO:BayesianNeuralIPS} is a \textcolor{red}{gradient-free and} sequential model-based method \textcolor{red}{to optimize expensive} black-box functions. In Bayesian optimization, the surrogate model (e.g. Gaussian process) and acquisition function (e.g. expected improvement) are applied to model the objective function and sample the point for evaluation in the next loop respectively. The method needs tested points to train the surrogate model. Thus the algorithm usually begins with a random search for the initial model building. In each loop after evaluation, the optimization algorithm generates a point by maximizing the acquisition function after updating the surrogate model. 
This paper uses Tree-structured Parzen estimator (TPE) \cite{BBO:TPE} to search multipliers, which is a variation of BO.

\section{\textcolor{black}{Proposed} Approximate Multiplier Design}\label{Sec:design}

\subsection{HA Array \textcolor{black}{Generation}}

\begin{figure}[ht]
  \begin{minipage}[t]{0.45\linewidth}
    \centering
    \includegraphics[width=\linewidth]{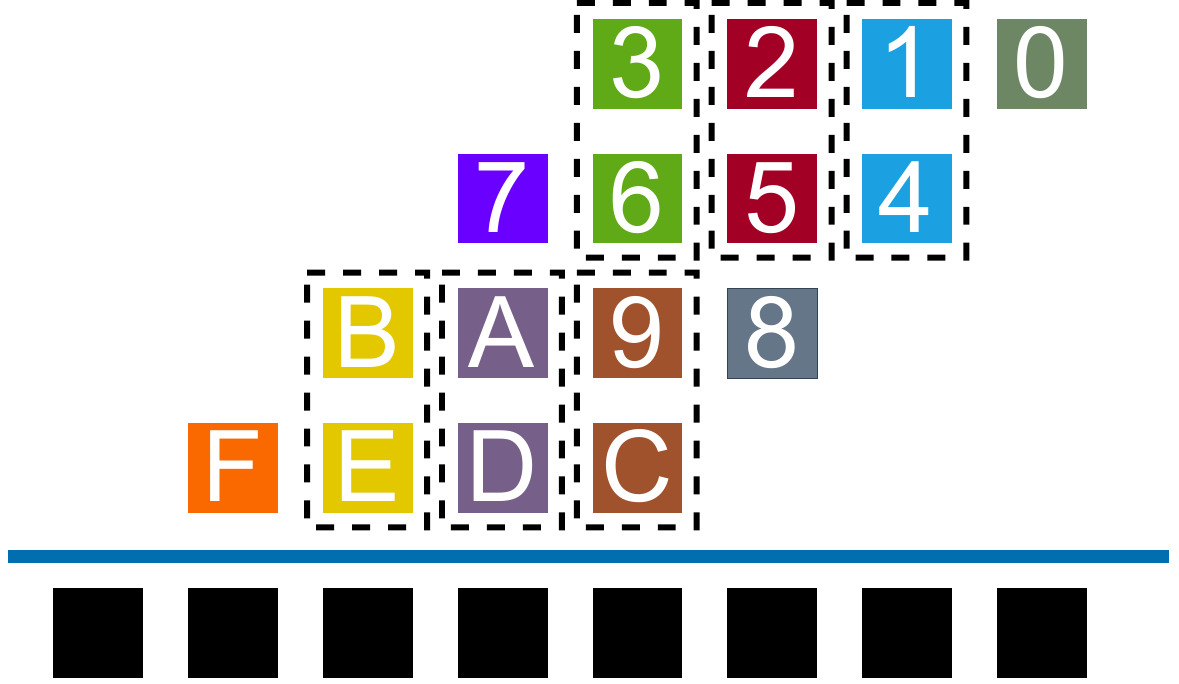}
    \caption{Uncompressed partial products of a 4$\times$4 unsigned multiplier.}
    \label{Fig:4x4}
  \end{minipage}
  \ \ \
  \begin{minipage}[t]{0.45\linewidth}
    \centering
    \includegraphics[width=\linewidth]{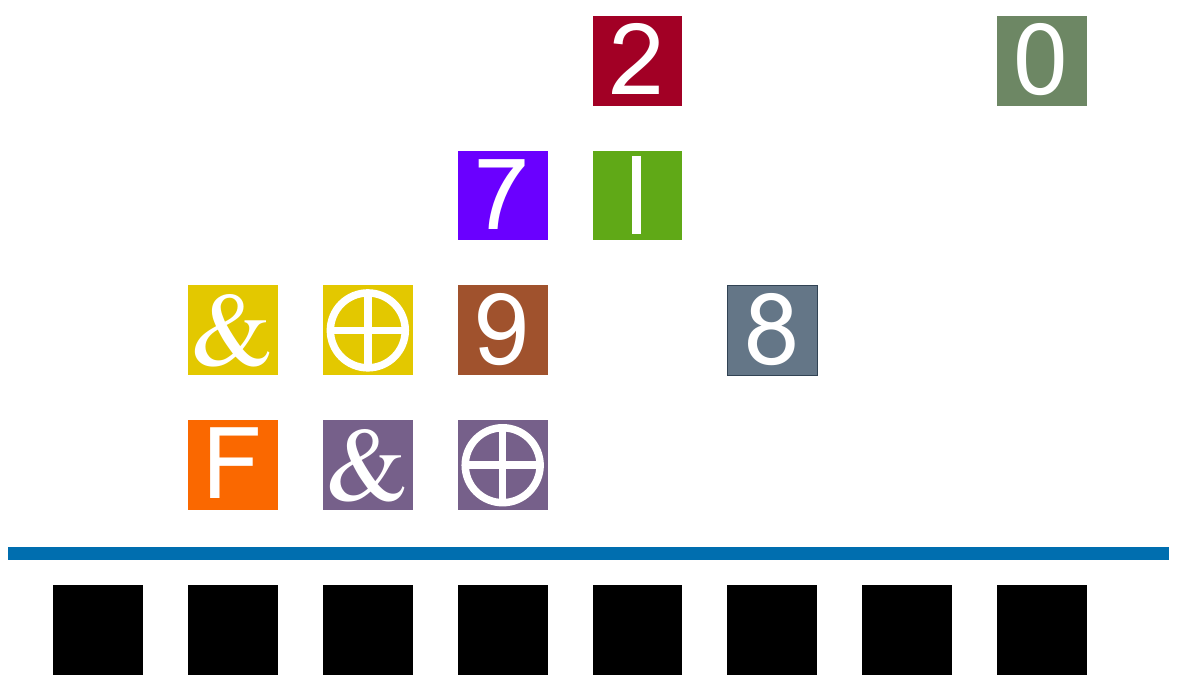}
    \caption{Compressed partial products based on optimized HAs of a 4$\times$4 unsigned multiplier.}
    \label{Fig:4x4_C}
  \end{minipage}
\end{figure}

Fig.~\ref{Fig:4x4} presents an uncompressed PP array of a 4$\times$4 unsigned multiplier \textcolor{black}{labeled with hexadecimal numbers}, which is produced by 16 AND gates. Dashed rectangles including two PPs represent exact HAs, forming a HA array. In Fig.~\ref{Fig:4x4}, a HA outputs a $Sum$ and a $Cout$. \textcolor{black}{All outputs of the HA array along with four uncompressed PPs including $\text{PP}_0$, $\text{PP}_7$, $\text{PP}_8$, and $\text{PP}_\text{F}$ construct a compressed PP array for the following accumulation.}
Assume that a multiplier has $N$ rows of PPs and each row contains $M$ PPs. The number of required exact HAs for the initial compression of an exact multiplier can be calculated by:
\begin{equation}
    \label{Eq:HA_number}
    S = (M-1) \times \lfloor \frac{N}{2} \rfloor
\end{equation}
where $\lfloor \ \rfloor$ means rounding down.
\textcolor{black}{For example, $S=6$ for a 4$\times$4 unsigned multiplier as shown in Fig.~\ref{Fig:4x4}.}
If we can simplify the circuit of the exact HA, the hardware cost of the multiplier will decrease.
It should be mentioned that the number of \textcolor{black}{uncompressed PPs is}:
\begin{equation}
    \label{Eq:Uncompress_number}
    N + (\textcolor{red}{N}\bmod 2) \times (M-1)
\end{equation}

\subsection{Simplification for An Exact HA}

We design an efficient method to optimize an exact HA, which has four options:
\begin{itemize}
    \item Eliminate: delete the HA straightforwardly, both $Sum$ and $Cout$ are connected to the ground.
    \item \textcolor{black}{OR $Sum$}: the output $Sum$ is ORed by HA inputs and the other output $Cout$ is connected to the ground.
    \item \textcolor{black}{Direct $Cout$}: \textcolor{black}{this option connects $Cout$ to one of the two inputs} and connects $Sum$ to the ground.
    \item Exact: keep the HA exact.
\end{itemize}
In mentioned options, ``Eliminate'' and ``OR $Sum$'' make the HA has a negative error. Conversely, ``Direct $Cout$'' makes the HA has a positive error. The proposed simplification method can improve the generated multiplier's precision by combining negative and positive errors.

\subsection{Precision Analysis}

We set a HA's weight to be the same as \textcolor{black}{the binary weight of its two inputs, which reflects the HA's significance to the final result. For example, in Fig.~\ref{Fig:4x4}, the weights of two HAs whose inputs are $\text{PP}_{1,4}$ and $\text{PP}_\text{B,E}$ are 1 and 5 respectively.}
After \textcolor{black}{the HA array generation}, we can use \textcolor{red}{BO} to \textcolor{black}{obtain} approximate multipliers.
Theoretically, since the search space of an HA contains the ``Exact'' option, the BO will tend to approximate low-weight HAs, keeping the high-weight HAs exact to generate high-precision multipliers with low hardware costs.
However, there is no such an actual efficient \textcolor{red}{BO implementation} that meets our requirements.
Thus we keep high-weight HAs accurate \textcolor{black}{manually} before \textcolor{red}{using BO}.
We assume that the area (i.e. LUT count) of a multiplier is \textcolor{black}{proportional to $S$ (see \eqref{Eq:HA_number})} and use \textcolor{black}{$R$} to represent the desired area percentage reduction.
The number of HAs to be searched is $\lfloor S \times R \rceil$ and the number of pre-reserved exact HAs is $ S - \lfloor S \times R \rceil$, where $\lfloor \ \rceil$ means rounding to nearest integer. 
All PPs compressed by the searched and pre-reserved exact HAs are summed up using Verilog ``$+$'' operators \textcolor{black}{along with original uncompressed PPs (e.g. $\text{PP}_\text{0,7,8,F}$ in Fig.~\ref{Fig:4x4})}.
It should be mentioned that the searched HAs may also be exact.
\textcolor{black}{Fig.~\ref{Fig:4x4_C} illustrates a compressed PP array based on optimized HAs with $R=0.8$. We can see that the total number of PPs after compression is 11, which is a 31.25\% reduction compared to Fig.~\ref{Fig:4x4}. Two HAs whose inputs are $\text{PP}_\text{B,E}$ and $\text{PP}_\text{A,D}$ are pre-reserved exact and search-based exact respectively.}

\subsection{Cost Function}

The $cost$ function is a critical element in \textcolor{red}{BO}, which determines the search direction. For example, if we set $cost$ as MAE, the \textcolor{red}{BO} algorithm will ignore hardware overhead and look for solutions that minimize MAE only. However, the quality of the approximate multiplier is jointly determined by hardware cost and error. Hence we need a unified metric to measure the quality of the approximate multiplier.
The hardware cost of the multiplier can be given by the PDA described in \textcolor{black}{Section} \ref{SubSec:PDA} and the error can be represented by the product of MAE and MSE (MM). A simple solution is to set the $cost$ to the product of PDA and MM. However, according to our analysis, MAE and MSE values of different approximate multipliers increase exponentially with the improvement of PDA, which means the $cost$ will be dominated by MM. In order to make multipliers with small $cost$ values located at the Pareto front, we set the $cost$ as PDAE, which is defined as:
\begin{equation}
    \label{Eq:PDAE}
    \text{PDAE} = \text{PDA} \times \log_2 (\text{MM}^{\prime})
\end{equation}
where
\begin{equation}
    \label{Eq:MMp}
    \text{MM}^{\prime} = \text{MAE} \times \text{MSE} + 1
\end{equation}
According to \eqref{Eq:PDAE} and \eqref{Eq:MMp}, the PDAE value of the exact multiplier is 0.

\subsection{Optimization Flow}

\begin{figure}[!htbp]
    \centering
    \includegraphics[width=0.9\linewidth]{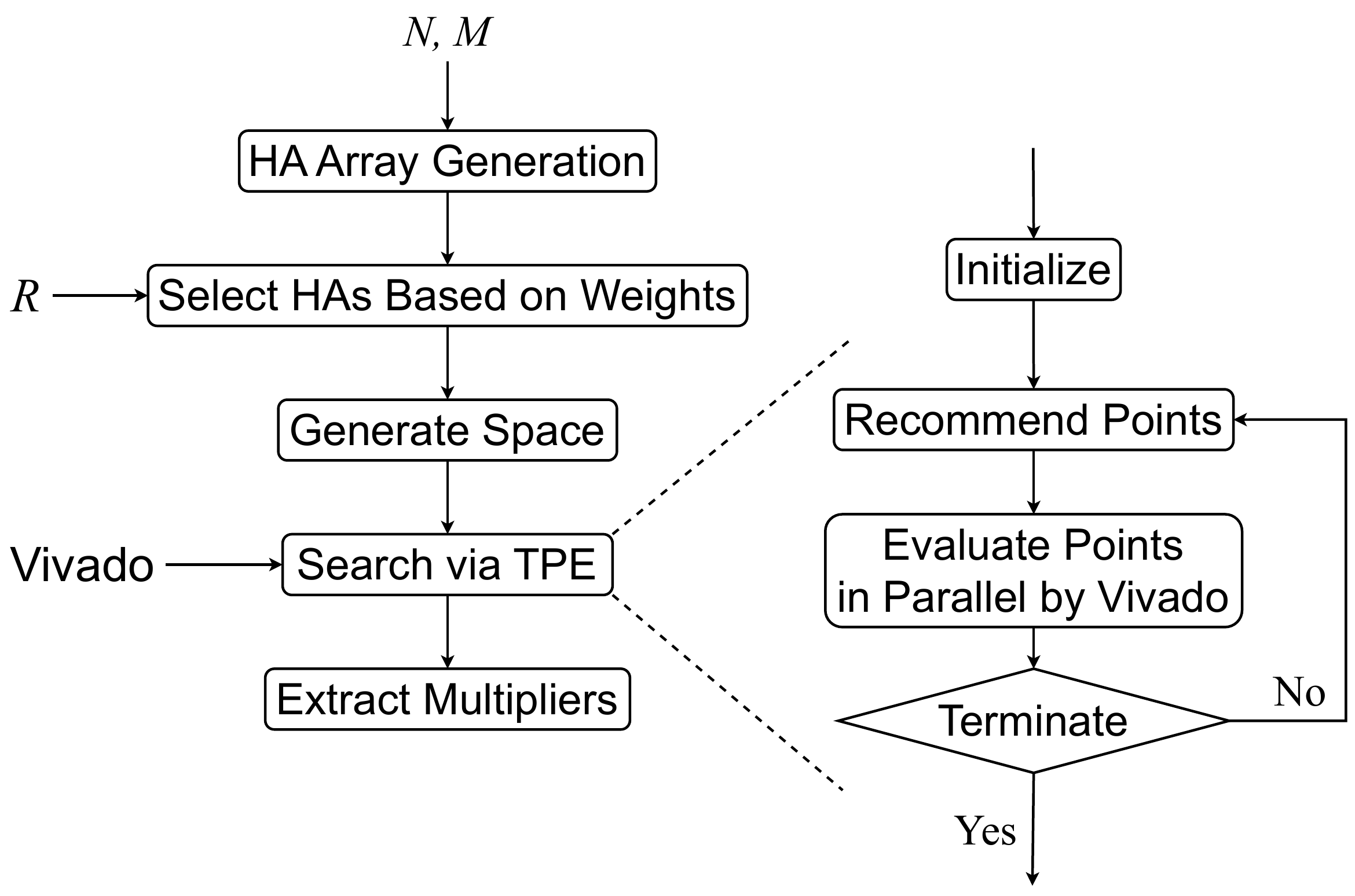}
    \caption{Optimization flow of the method.}
    \label{Fig:flow}
\end{figure}

Fig.~\ref{Fig:flow} shows the optimization flow of our proposed method. Firstly, the HA array is produced with the given multiplier's bit widths $N$ and $M$. Secondly, we select $\lceil S \times R \rceil$ HAs with the smallest weights to generate a search space.
\textcolor{black}{Then we use TPE algorithm to recommend points after the initialization.} For each point needed to be evaluated, \textcolor{black}{Vivado is applied} to simulate, synthesize, place and route to \textcolor{black}{obtain} the $cost$ value. \textcolor{black}{When the termination condition is reached,} we extract multipliers that are on the Pareto front.

\section{Experimental Results}\label{Sec:results}

\subsection{Setup}

\textcolor{black}{To comprehensively evaluate the proposed automated method,} we set the the desired area percentage reduction $R=0.3, \ 0.4, \ 0.5, \ 0.6, \ 0.7$ with the unsigned 8$\times$8 multiplier and run the search process for 48 hours on a 60-core Intel Xeon server independently. 
After that, we extract multipliers that are on the Pareto front.

\textcolor{black}{For a fair comparison} of the performance of the generated unsigned 8$\times$8 multipliers, we simulate and implement many approximate multipliers from previous works to \textcolor{black}{obtain} their hardware costs and errors, which includes SDLC \cite{Approximate:SDLC}, KMap \cite{Approximate:KMap}, AC \cite{Approximate:AC}, RoBA \cite{Approximate:RoBA}, CR \cite{Approximate:CR}, OU \cite{Approximate:Optimal}, DRUM \cite{Approximate:DRUM}, TOSAM \cite{Approximate:TOSAM}, PPAM \cite{Approximate:PPAM}, EvoApprox8b \cite{Approximate:EvoApprox8b}, EvoApproxLite \cite{Approximate:EvoLite}, FPT22 \cite{Approximate:FPT22}, \textcolor{red}{CaCc\cite{Approximate:CaCc}}, SMApproxLib \cite{Approximate:SMApproxLib}, ApproxFPGAs \cite{Approximate:ApproxFPGAs}, and TCAD22 \cite{Approximate:TCAD22}. SDLC is implemented with 2-bit depth compression, which achieves the highest precision. Error recoveries in CR multipliers are configured with 6 bits and 7 bits. The floating-point multiplier OU is reproduced to an unsigned integer multiplier with level-1 error compensation. DRUM is designed with $m = 4, 5, 6, 7$. TOSAM is reproduced with $(h, t) \in \{1, 2, 3\} \times \{3, 4, 5, 6, 7\}$. PPAM can be specified by the user to reduce errors with or without comparators. We reproduce it by $(j, k) \in \{0, 1, 2\} \times \{1, 2, 3\}$ with both versions. Besides, two exact multipliers and truncation-based approximate multipliers are also compared.

We use Synopsys VCS S-2021.09-SP2 and Vivado 2023.1 to simulate, synthesize, place, and route multipliers with Xilinx Virtex UltraScale+ device xcvu3p-ffvc1517-3-e.

\subsection{Result}

\subsubsection{Scatter Plot}

\begin{figure*}[!htbp]
    \centering
    \includegraphics[width=\linewidth]{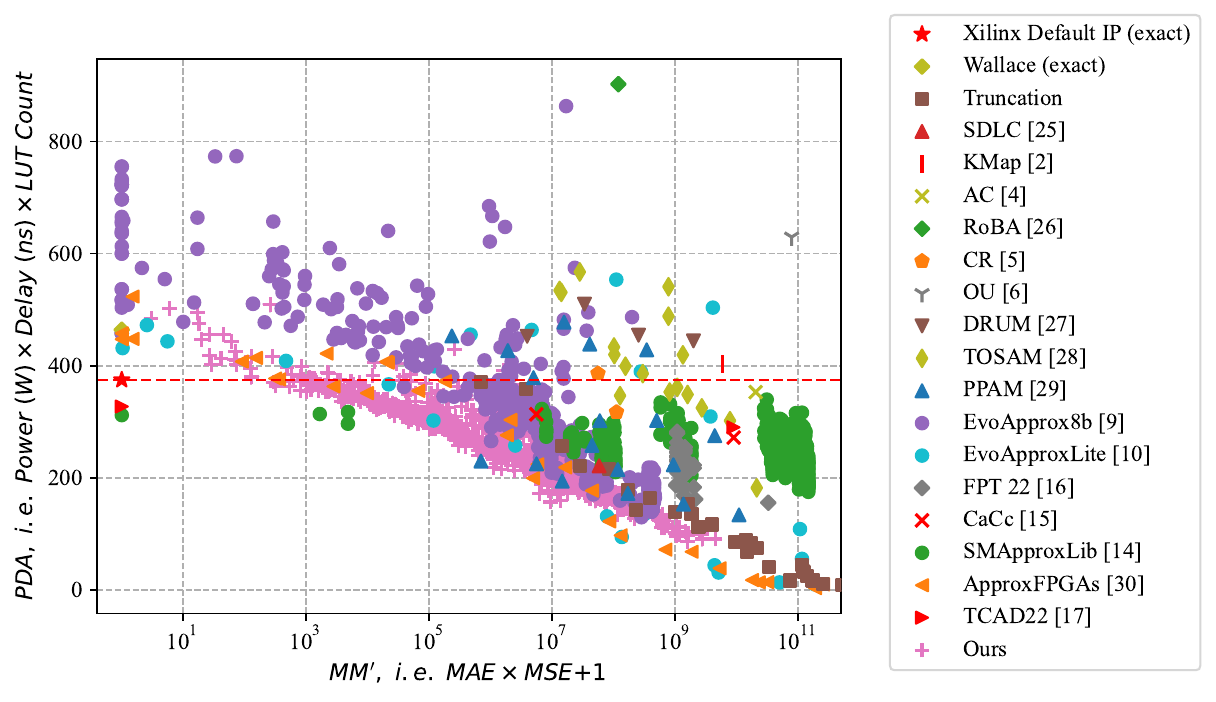}
    \caption{Scatter plot of PDA and MM$^{\prime}$ based on different multipliers.}
    \label{Fig:compare}
\end{figure*}

\begin{table*}[!htbp]
    \renewcommand{\arraystretch}{1.4}
    \caption{Best PDAE Value of different Multipliers Group}
    \begin{center}
        \begin{tabular}{|c|c|c|c|c|c|c|c|c|}
        \hline
         & \multicolumn{2}{|c|}{\textbf{$\text{MM}^{\prime} \in [10^3,10^7] $}} & \multicolumn{2}{|c|}{\textbf{$\text{MM}^{\prime} \in [10^3,10^8] $}} & \multicolumn{2}{|c|}{\textbf{$\text{MM}^{\prime} \in [10^4,10^7] $}} & \multicolumn{2}{|c|}{\textbf{$\text{MM}^{\prime} \in [10^4,10^8] $}} \\
        \cline{1-9}
        \textbf{\textit{Group}} & \textbf{\textit{Best PDAE}} & \textbf{\textit{Imp. (\%)}} & \textbf{\textit{Best PDAE}} & \textbf{\textit{Imp. (\%)}} & \textbf{\textit{Best PDAE}} & \textbf{\textit{Imp. (\%)}} & \textbf{\textit{Best PDAE}} & \textbf{\textit{Imp. (\%)}}  \\
        \hline
        Truncation & 7205.03 & 50.90 & 5488.09 & 35.53 & 7205.03 & 49.60 & 5488.09 & 33.83 \\
        \hline
        SDLC \cite{Approximate:SDLC} & \multirow{6}{*}{5709.04} & \multirow{6}{*}{38.03} & \multirow{6}{*}{5709.04} & \multirow{6}{*}{38.03} & \multirow{6}{*}{5709.04} & \multirow{6}{*}{36.39} & \multirow{6}{*}{5709.04} & \multirow{6}{*}{36.39} \\
        KMap \cite{Approximate:KMap} & & & & & & & & \\
        AC \cite{Approximate:AC} & & & & & & & & \\
        RoBA \cite{Approximate:RoBA} & & & & & & & & \\
        CR \cite{Approximate:CR} & & & & & & & & \\
        OU \cite{Approximate:Optimal} & & & & & & & & \\
        \hline
        DRUM \cite{Approximate:DRUM} & 9909.35 & 64.30 & 9909.35 & 64.30 & 9909.35 & 63.35 & 9909.35 & 63.35 \\
        \hline
        TOSAM \cite{Approximate:TOSAM} & 6247.27 & 43.37 & 6247.27 & 43.37 & 6247.27 & 41.87 & 6247.27 & 41.87 \\
        \hline
        PPAM \cite{Approximate:PPAM} & 4461.53 & 20.70 & 4461.53 & 20.70 & 4461.53 & 18.61 & 4461.53 & 18.61 \\
        \hline
        EvoApprox8b \cite{Approximate:EvoApprox8b} & 4857.81 & 27.17 & 4340.22 & 18.48 & 4857.81 & 25.25 & 4340.22 & 16.33 \\
        \hline
        EvoApproxLite \cite{Approximate:EvoLite} & 5088.03 & 30.46 & 3442.06 & -2.79 & 5088.03 & 28.63 & 3442.06 & -5.50 \\
        \hline
        FPT22 \cite{Approximate:FPT22} & 5010.96 & 29.40 & 5010.96 & 29.40 & 5010.96 & 27.53 & 5010.96 & 27.53 \\
        \hline
        CaCc \cite{Approximate:CaCc} & 7017.36 & 49.58 & 7017.36 & 49.58 & 7017.36 & 48.25 & 7017.36 & 48.25 \\
        \hline
        SMApproxLib \cite{Approximate:SMApproxLib} & 3356.49 & -5.41 & 3356.49 & -5.41 & 6266.03 & 42.05 & 5801.66 & 37.41 \\
        \hline
        ApproxFPGAs \cite{Approximate:ApproxFPGAs} & 4151.85 & 14.79 & 3220.87 & -9.85 & 4429.49 & 18.02 & 3220.87 & -12.74 \\
        \hline
        TCAD22 \cite{Approximate:TCAD22} & 9584.22 & 63.09 & 9584.22 & 63.09 & 9584.22 & 62.11 & 9584.22 & 62.11 \\
        \hline
        Ours & 3537.97 & / & 3537.97 & / & 3631.35 & / & 3631.35 & / \\
        \hline
        Avg. Imp. & / & 35.53 & / & 28.70 & / & 38.47 & 30.62 & / \\
        \hline
        \end{tabular}
        \label{Tab:PDAE}
        \end{center}
\end{table*}

We calculate PDA and MM$^{\prime}$ values of different multipliers and draw a scatter plot as shown in Fig.~\ref{Fig:compare}. We can see that our multipliers form a Pareto front. Two multipliers from SMApproxLib and TCAD22 are exact multipliers with lower PDA values than the exact multiplier built from Xilinx default IP. After checking the original data, we found that these two multipliers consume $\sim$20\% fewer LUTs than the Xilinx IP with a cost of $\sim$5\% increase in critical path delay, which is consistent with the result presented in \cite{Approximate:SMApproxLib} and \cite{Approximate:TCAD22}. Three multipliers from SMApproxLib based on the area-oriented exact multipliers achieve good results with small errors (MM$^{\prime} \in [10^3,10^4]$).
Multipliers from FPT22, CaCc, SMApproxLib, and TCAD22 libraries are designed with the LUT encoding method. According to Fig.~\ref{Fig:compare}, these manual-design multipliers have large error differences.
In contrast, multipliers based on automated approaches such as EvoApprox8b and ours have small performance differences, which can provide more options for various applications with different constraints.
The last observation is that EvoApproxLite and ApproxFPGAs multipliers have better performance than others when MM$^{\prime} \in [10^8,10^{11}]$. But these multipliers may not be practical because they require applications to have a large error tolerance.

\subsubsection{Effectiveness of Cost Function}

Most of the multipliers with small PDAE values extracted from the set of generated multipliers are on Pareto front, which demonstrates the effectiveness of the proposed PDAE $cost$ function.

\subsubsection{PDAE Comparison}

In order to compare the performance of different multipliers intuitively, we group the multipliers and calculate the best PDAE of each \textcolor{black}{group} with four ranges of MM$^{\prime}$.
Table \ref{Tab:PDAE} presents the best PDAE values of different multiplier groups.
Our multipliers achieve 28.70\%-38.47\% improvements on average.
It should be mentioned that if we limit the MM$^{\prime}$ with the range of $[10^4,10^7]$, our generated multipliers have the best performance among all counterparts.

\section{Conclusion}\label{Sec:conclusion}

In this paper, we propose \textcolor{red}{AMG}, an automated approximate multipliers generator by \textcolor{red}{Bayesian} optimization for FPGA-based systems, which can parallelly search the optimized HA array for the partial product compression. The generated multiplier can be efficiently mapped to lookup tables and fast carry chains to reduce the hardware overhead. Extensive experimental results show that our multipliers can form a Pareto front with up to a 38.47\% improvement on average than previous works. Although the experiment is carried out based on FPGAs and uniform distributions, the method can be easily extended to ASICs and non-uniform distributions.



\bibliographystyle{IEEEtran}
\bibliography{IEEEexample.bib}

\end{document}